\documentclass[prl,aps,twocolumn,showpacs]{revtex4}

\usepackage{epsfig}

\begin{document}

\title{Quenched Fe Moment in the Collapsed Tetragonal Phase of Ca$_{1-x}$Pr$_{x}$Fe$_2$As$_2$}

\author{Long Ma$^{1}$}
\author{G. F. Ji$^{1}$}
\author{J. Dai$^{1}$}
\author{S. R. Saha$^{2}$}
\author{J. Paglione$^{2}$}
\author{Weiqiang Yu$^{1}$}
\email{wqyu_phy@ruc.edu.cn}
\affiliation{
$^{1}$ Department of Physics, Renmin University of China, Beijing 100872, China\\
$^{2}$ Center for Nanophysics and Advanced Materials, Department of Physics, University of Maryland, College Park, MD 20742, USA}
\date{\today}

\begin{abstract}

We report $^{75}$As NMR studies on single crystals of rare-earth doped iron pnictides superconductor Ca$_{1-x}$Pr$_{x}$Fe$_{2}$As$_{2}$ ($x$=0.075 and 0.15). The $^{75}$As spectra show a chemical 
pressure effect with doping and a first order structure transition to the collapsed tetragonal phase upon cooling. A sharp drop of the Knight shift is seen below the structural transition, whereas 
$1/T_1$ is strongly enhanced at low-temperatures. These evidences indicate quenching of Fe local magnetism and short-range ordering of Pr$^{3+}$ moment in the collapsed tetragonal phase. The quenched 
Fe moment through structure collapse suggests a strong interplay of structure and magnetism, which is important for understanding the nature of the collapsed tetragonal phase.

\end{abstract}

\maketitle

The interplay of structure, magnetism, and superconductivity is one of the most essential ingredients to understand the high-temperature superconductivity in the iron-based materials. At low 
temperatures, the occurrence of superconductivity is triggered by the suppression of the antiferromagnetic, orthorhombic ($\mathcal{O}$) phase to the paramagnetic, tetragonal ($\mathcal{T}$) phase 
either by carrier doping or by pressure \cite{Hosono_Jacs_130_3296, Chen_Nature_453_761, Ren_CPL_12_105, Chen_PRL_100_247002, Rotter_PRL_101_107006}. Recently, a high-temperature $\mathcal{T}$ to a 
low-temperature collapsed tetragonal ($c\mathcal{T}$) phase without magnetic ordering has also been reported by pressure or doping in many iron pnictides, such as CaFe$_2$As$_2$ ($P$$\ge$0.3 GPa) 
\cite{Torikachvili_PRL_101_057006, Kreyssig_PRB_78_184517, Park_CM_20_322204}, rare-earth doped Ca$_{1-x}$$R_x$Fe$_2$As$_2$ ($P$$=$0, $R$=Pr, Nd) \cite{Saha_arxiv_1105_4798, Chu_PNAS}, BaFe$_2$As$_2$ 
($P$$\ge$27 GPa)\cite{Mittal_PRB}, EuFe$_2$As$_2$($P$$\ge$8 GPa) \cite{Uhoya}, CaFe$_2$ (As$_{1-x}$P$_x$)$_{2}$ ($P$$=$0) \cite{Kasahara_PRB_83_060505}, and Fe$_{1.05}$Te \cite{Sun_PRB_80_114519}. 
Interestingly, superconductivity is found in the first two compounds accompanying the emergence of the $c\mathcal{T}$ phase at low temperatures, but with a small volume ratio 
\cite{Torikachvili_PRL_101_057006, Park_CM_20_322204, Saha_arxiv_1105_4798, Chu_PNAS}. CaFe$_2$As$_2$ has a $T_C$ about 12 K in liquid pressure cell \cite{Kreyssig_PRB_78_184517, Park_CM_20_322204}, 
but loses superconductivity in the pure $c\mathcal{T}$ phase under hydrostatic condition \cite{Yu_PRB_79_020511}. NMR studies show superconductivity in the residual $\mathcal{T}$ structure 
\cite{Baek_PRL, Kawasaki_super_23_054004}, indicating that superconductivity occurs in the interface between $\mathcal{T}$-strcture and $c\mathcal{T}$-structure domains due to strain 
effect\cite{Yu_PRB_79_020511}. 

For rare-earth doped Ca$_{1-x}$$R_x$Fe$_2$As$_2$, double superconducting transition with $T_C$ about 10 K and 45-47 K have been reported \cite{Saha_arxiv_1105_4798, Chu_PNAS}. The low $T_C$ phase is 
probably an analogy to the strain effect in CaFe$_2$As$_2$ \cite{Chu_PNAS}. The nature of the high-$T_C$ phase remains unclear, since its $T_C$ is much higher than the electron-doped 
Ca(Fe$_{1-x}$Co$_x$)$_2$As$_2$ \cite{Harnagea_arxiv_1011_2085}, as well highest among the pnictides with the ThCr$_2$Si$_2$ (122) structure. In order to understand the high-$T_C$ phase, it is 
essential to ask whether spin fluctuations exist in the $c\mathcal{T}$ of the rare-earth doped CaFe$_2$As$_2$. The structure collapse is usually induced by a large reduction of the $c$-axis lattice 
parameter (or the $c/a$ ratio) by external or chemical pressure \cite{Kreyssig_PRB_78_184517, Goldman_arxiv.0811.2013,Saha_arxiv_1105_4798}. For CaFe$_2$As$_2$, high-pressure inelastic neutron 
scattering (INS) indicates that dynamic magnetic correlations are suppressed at the AFM wave vector ($\frac{\pi}{a}$,$\frac{\pi}{a}$) in the high-pressure $c\mathcal{T}$ phase of CaFe$_2$As$_2$ 
\cite{Pratt_prb_79_060510}, but spin fluctuations in other momentum modes are not investigated. Theoretical studies in the 122 structure suggests two scenarios, the reduced As-As interlayer distance 
under pressure may lead to a change of Fermi surface topology and therefore suppresses the Fe moment either in the $c\mathcal{T}$ phase \cite{Yildirim_PRL_102_037003, Kasahara_PRB_83_060505, 
ZhangYZ_prb_80_094530}, or already in the high-temperature $\mathcal{T}$ phase \cite{Ji_PRB_83_132504}.  For Ca$_{1-x}$$R_x$Fe$_2$As$_2$, it is not clear if the magnetic correlations are 
suppressed/enhanced upon rare-earth doping in the $\mathcal{T}$ phase and the $c\mathcal{T}$ phase. 

In this report, we present $^{75}$As ($S=3/2$) NMR studies on Ca$_{1-x}$Pr$_{x}$Fe$_2$As$_2$ ($x$$=$0.075, 0.15) superconducting single crystals. We first find NMR evidence for structure collapse from 
the $^{75}$As quadrupole effect. We then study the magnetic properties of the $c\mathcal{T}$ phase from both the Knight shift and the spin-lattice relaxation rate. Our Knight shift $^{75}K$ drops 
sharply below the structure transition $T_S$, which indicated suppressed Fe paramagnetic moments in the $c\mathcal{T}$ phase. The spin-lattice relaxation rate $1/^{75}T_1$, on the other hand, is 
strongly enhanced below $T_S$ and forms a broad peak at $T^*$ $\sim$ 25 K. The contrasting behavior of $^{75}K$ and $1/^{75}T_1$ suggests a short-range ordering from Pr$^{3+}$. Such magnetic 
properties are not only necessary for understanding the physical nature of structure collapse, but also provide important information for unveiling the high-$T_C$ phase.

\begin{figure}
\includegraphics[width=8.5cm, height=6cm]{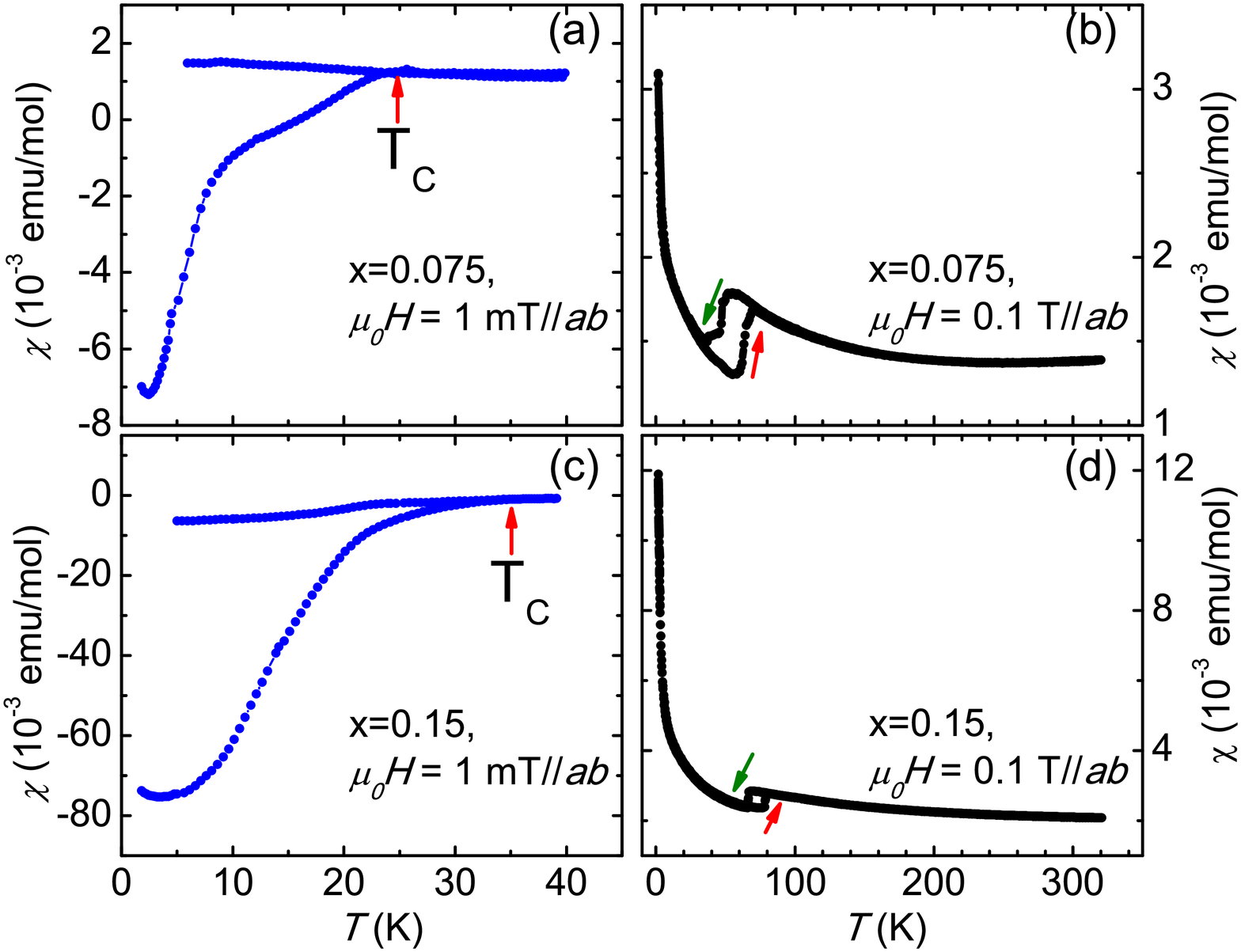}
\caption{\label{MT}(color online) dc susceptibility as a function of temperature for the $x$=0.075 and $x$=0.15 samples respectively. (a) and (c) shows the susceptibility under a magnetic field of 1 
mT, and (b) and (d) under a higher field of 0.1 T, with fields applied along the crystalline $ab$-plane.}
\end{figure}

\begin{figure}
\includegraphics[width=8.5cm, height=5cm]{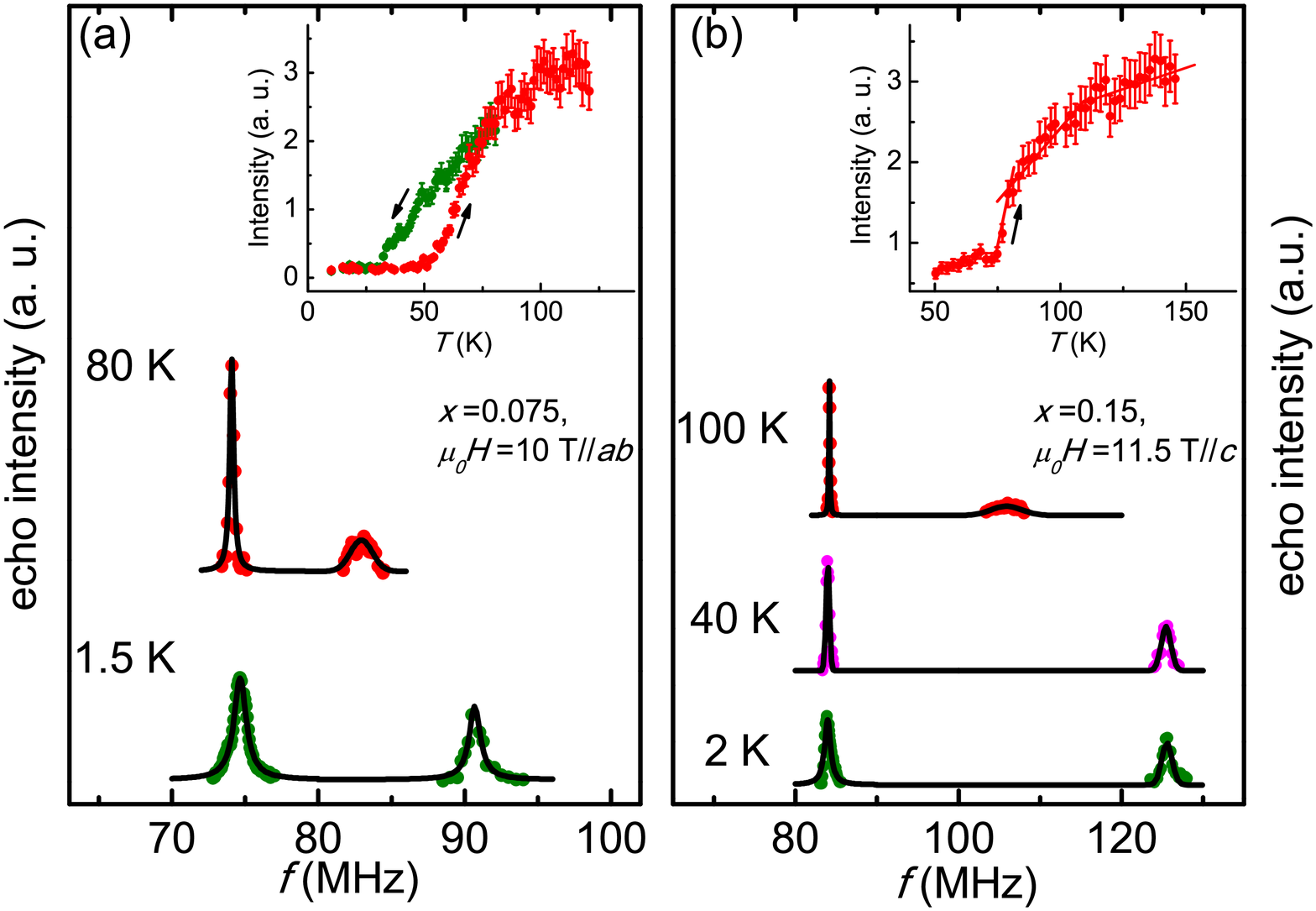}
\caption{\label{struc}(color online) The $^{75}$As NMR spectra at typical temperatures for (a) the $x$$=$0.075 crystal with a 10 T field applied in the crystalline $ab$-plane and (b) the $x$$=$0.15 
crystal with an 11.5 T field applied along the $c$-axis. Inset: the temperature dependence of the spectral weight at the frequency of the central transition of the $\mathcal{T}$ phase, with arrows 
indicating the warming (cooling) direction. }
\end{figure}

The Ca$_{1-x}$Pr$_{x}$Fe$_{2}$As$_{2}$ single crystals were synthesized by FeAs self-flux method with details reported previously \cite{Saha_arxiv_1105_4798}. For our NMR experiments, single crystals 
with area of $\sim$5$\times$5 mm$^2$ and thickness of 0.5 mm along crystalline $c$-axis were chosen. We studied samples with two compositions with $x$$=$0.075 and 0.15 (determined by WDS), which were 
characterized by the susceptibility measurements in SQUID, as shown in Fig.~\ref{MT}. The NMR experiments are carried out using the standard coherent pulse method. The sample is placed on a rotator to 
allow to change field orientation, and the frequency-swept NMR spectra are obtained by integrating the intensity of Fourier transform of the spin echo signal. The spin-lattice relaxation rate $1/T_1$ 
is obtained by the inversion-recovery method. Most data are collected during warming up process for consistency, except the hysteresis loop in Fig.~\ref{struc}(a). 

For two doping levels, both the superconducting transition and the structure transition are evident from the susceptibility data. As shown in Fig.~\ref{MT}(a), with a 1 mT in-plane field, the 
$x$=0.075 single crystal superconducts below $T_c\sim$ 25 K, and superconducting volume ratio is estimated to be about 0.2$\%$. The $x$=0.15 single crystal has a higher $T_c \sim$35 K and a larger 
superconducting volume $\sim$ 2\% as shown in Fig.~\ref{MT}(c). With a 0.1 T magnetic field applied in the $ab$-plane of the single crystals, the superconductivity of both crystals is suppressed, as 
shown in Fig.~\ref{MT}(b) and (d). Such a small volume of superconductivity is consistent with previous reports \cite{Saha_arxiv_1105_4798, Chu_PNAS}. Furthermore, the susceptibility under 0.1 T field 
shows a clear hysteresis loop, corresponding to the first-order, $\mathcal{T}$ to $c\mathcal{T}$ phase structural transition \cite{Saha_arxiv_1105_4798}. The transition temperature $T_S$ is about 64 K 
(warming) and 48 K (cooling) for the $x$=0.075 sample, and 78 K (warming) and 66 K (cooling) for the $x$=0.15 sample. 

The $^{75}$As NMR spectra at typical temperatures are shown in Fig.~\ref{struc}(a) and (b) for two crystals. We note here that superconductivity is already suppressed by the large NMR field. 
Fig.~\ref{struc}(a) shows the $^{75}$As NMR spectra for the $x$$=$0.075 crystal, with a central transition ($f\approx$74.5 MHz) and a high-frequency satellite ($f\approx$83 MHz above $T_S$ and 
$f\approx$90.7 MHz below $T_S$), with a 10 T field applied in the $ab$ plane. Similar spectra for the $x$$=$0.15 crystal are shown in Fig.~\ref{struc}(b) with a 11.5 T field applied along the 
$c$-axis.

We first look at the $\mathcal{T}$ phase ($T$$>$$T_S$).  Since the As site has a local fourfold in-plane symmetry, the principal axis of the electric field gradient is along the crystalline $c$-axis. 
The nuclear quadrupole resonance frequencies ($\nu_q$) can be estimated by the angular dependence of the satellite frequency, $f=\nu_L(1+K)\pm\nu_q(3cos^2\theta-1)/2$, where $\nu_L$ and $K$ represent 
the Larmor frequency and the Knight shift respectively, and $\theta$ is the angle between the magnetic field and the crystalline $c$-axis. Since $K$ is small (shown later), the $\nu_q$ above T$_S$ is 
estimated to be $\sim$20.4 MHz for x=0.075, and  $\sim$21.6 MHz for x=0.15. For comparison, $\nu_q$ for the $\mathcal{T}$ phase of CaFe$_2$As$_2$ is about 11.8 MHz at $P$$=$0 and 25 MHz at P=1.08 GPa 
\cite{Kawasaki_super_23_054004}. The increase of $\nu_q$ with pressure is consistent with the increased local EFG field from reduced $c$-axis lattice parameters without crystal symmetry change. In 
contrast, the $\nu_q$ of the electron-doped Ca(Fe$_{1-x}$Co$_x$)$_2$As$_2$ barely change with doping \cite{Baek_PRB_79_052504}, which suggests a negligible contribution from carrier (electron) doping 
effect. Therefore, the large increase of $\nu_q$ in Ca$_{1-x}$Pr$_{x}$Fe$_{2}$As$_{2}$ suggests Pr$^{3+}$ doping produces a strong chemical pressure effect along the $c$-axis and induces the observed 
structure collapse at low temperatures.  

The structure transition is indicated by the shift of the high-frequency satellite through $T_S$, as shown in Fig.~\ref{struc}(a) and (b). The 
$\nu_q$ below $T_S$ are estimated to be 35.8 MHz for $x$$=$0.075, and 41.5 MHz for $x$$=$0.15. The large increase of the $\nu_q$ below $T_S$ of both dopings is again consistent with reduced $c$-axis 
lattice parameter without crystal symmetry change. The frequency of the central transition also varies with temperature. The angular dependence of the central frequency varies with a combination of 
Knight shift and a second-order quadrupole correction $ f=(1+K)\nu_L + \frac {-3\nu_Q^2}{16\nu_L}(1-cos^2\theta)(9cos^2\theta-1)$. The inset of Fig.~\ref{struc}(a) and (b) shows the spectral weight, 
at one frequency fixed at the central transition of the $\mathcal{T}$ phase, as a function of temperature. The sharp drop of the spectral weight upon structure collapse and the its thermal hysteresis 
loop are consistent with the susceptibility data as shown in the inset of Fig.~\ref{struc}(a) and (b).

\begin{figure}
\includegraphics[width=8.5cm, height=7cm]{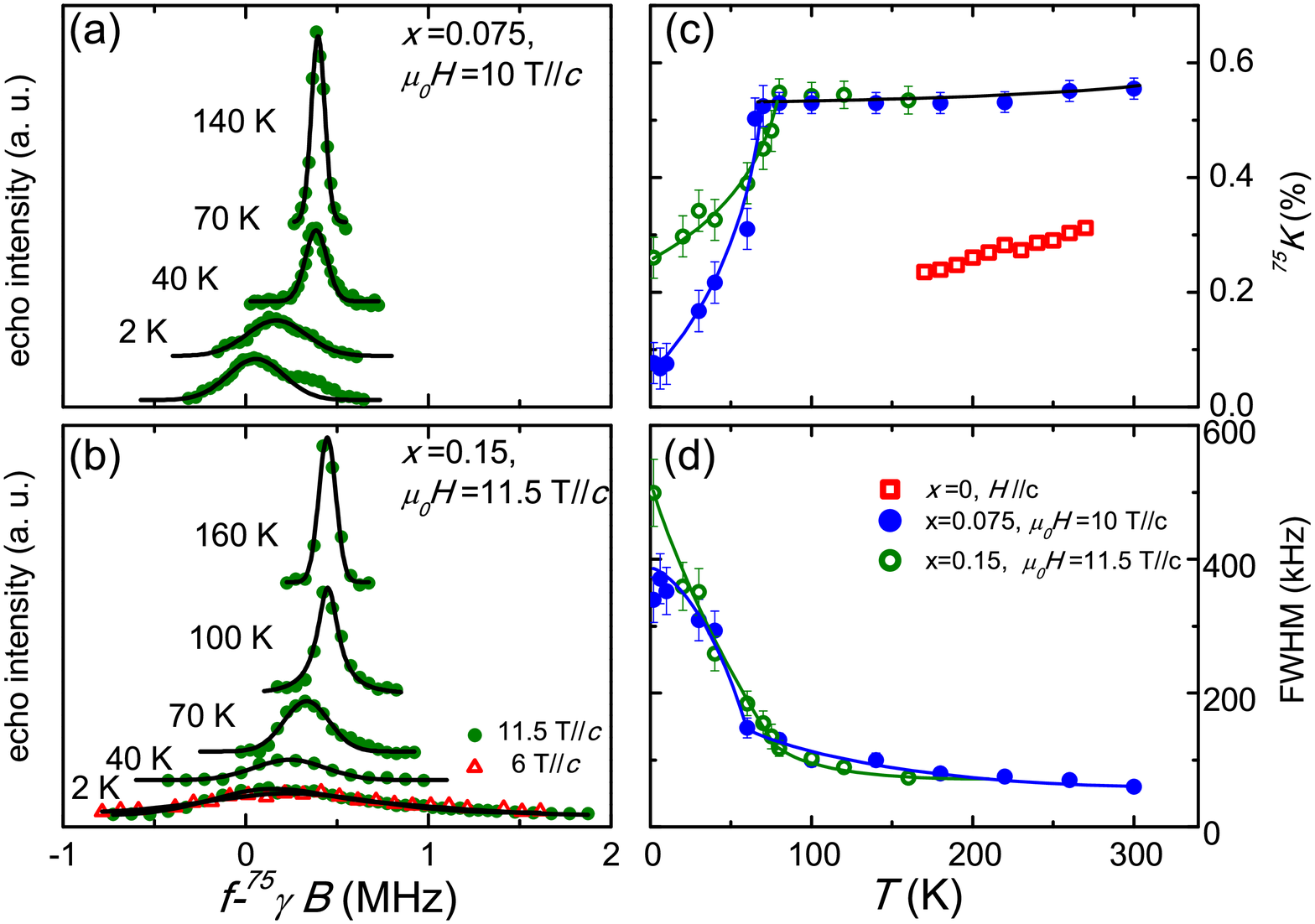}
\caption{\label{ks}(color online) (a) The $^{75}$As spectra of the central transition of the $x$$=$0.075 sample at typical temperatures with 10 T field along the $c$-axis. (b) The $^{75}$As spectra of 
the central transition of the $x$=0.15 sample at typical temperatures with 6 T and 11.5 T field along the $c$-axis. (b) The $^{75}$As Knight shifts of  Ca$_{1-x}$Pr$_{x}$Fe$_2$As$_2$ ($x$$=$0, 0.075, 
0.15) as a function of temperature. The $x$=0 data is adapted from Ref.~\onlinecite{Baek_PRB_79_052504}. (d) The FWHM of the $^{75}$As central transition as a a function of temperature for two 
dopings.}
\end{figure}
 
With field applied along the $c$-axis ($\theta$=0), the second order quadrupole correction to the central frequency is zero. Then the Knight shift can be calculated from the central frequencies by 
$^{75}K(T)=(f-^{75}\gamma B)/^{75}\gamma B$, where $^{75}\gamma$$=$7.292 MHz/T is the gyromagnetic ratio of $^{75}$As, and $f$ is the resonance frequency under field $B$. Fig.~\ref{ks}(a) and (b) 
shows the central transition for two dopings at typical temperatures with field applied along the c-axis. With reducing temperature, the central frequency shifts to low frequency and the spectra 
become broadened significantly.  The Knight shift and the NMR linewidth are obtained from the Gaussian fit to the spectra and depicted as a function of temperature as shown in Fig.~\ref{ks}(c) and 
(d). 

Above $T_S$, the Knight shift barely changes with temperature for both dopings, as shown in Fig.~\ref{ks}(c). The Knight shift of the undoped CaFe$_2$As$_2$ \cite{Baek_PRB_79_052504} is also plotted 
for comparison. The $K(T)$ increase from 0.32$\%$ to 0.58$\%$ with Pr$^{3+}$ doping from 0 to 0.075 at the ambient temperature. In other electron-doped iron pnictides, such as LaFeAsO$_{1-x}$F$_x$ 
\cite{Klingeler_PRB_81_024506}, LaFeAsO$_{1-\delta}$\cite{Mukuda_JPSJ_78}, and Ba(Fe$_{1-x}$Co$_x$)$_2$As$_2$\cite{Ning_PRL_104}, the Knight shift decreases slightly with electron doping. We think the 
increase of the high-temperature Knight shift with rare-earth doping is likely caused by the increased hyperfine coupling between $^{75}$As and Fe moments from the shrinkage of the $c$-axis lattice 
parameter.   

Fig.~\ref{ks}(c) shows that $^{75}K(T)$ decrease slightly in the $\mathcal{T}$ phase, and then drops sharply below $T_S$. We point out here that the drop of $^{75}K(T)$ through the structure collapse 
is not a Pauli paramagnetic effect. In fact, all transport measurements suggest that the carrier density on the Fermi surface is either unchanged or increased when cooled through the structure 
collapse \cite{Saha_arxiv_1105_4798}. With lattice shrinkage along the $c$-axis, Pauli paramagnetism should induce an increase of the $^{75}K$ through structure collapse. Therefore, the reduction of 
$^{75}K(T)$ suggests a spin correlation effect from reduced Fe moment. Since the NMR Knight shift $K$$\sim$$\chi_0$, where $\chi_0$ is the local susceptibility of electrons, the reduced $K(T)$ 
signifies a large suppression of local paramagnetic spin fluctuations from Fe upon structure collapse.  In particular $K(T)\approx$ 0 at $T=$2 K, which suggest that the Fe moment is almost zero in the 
$c\mathcal{T}$ phase. Similar reduction of $^{75}K(T)$ is obtained for $x=$ 0.15 below $T_S$ as shown in Fig.~\ref{ks}(c), although the reduction is not as dramatic as that of the $x$$=$0.075 sample.

However, the linewidth of the $^{75}$As central transition broadens at very low temperature ($T$$\le$$T_S$) for both dopings, as shown in Fig.~\ref{ks}(d). For the $x$$=$0.15 sample, it increases from 
100 KHz to 500 KHz monotonically with temperature from $T_S$ down to 2 K. Such broadening suggests an inhomogeneous electronic environment. We further checked the field dependence of the  $^{75}$As  
spectra at the lowest temperature $T$$=$2 K for the $x$$=$0.15 sample, as shown in Fig.~\ref{ks}(b). The NMR linewidth, about 500 KHz, does not change much with field increased from 6.5T to 11.5T. 
Such a broadening is not consistent with either a quadrupole effect ($\Delta f\propto$$1/H$) or an inhomogeneous Knight shift ($\Delta f\propto$$H$), but points a local magnetic ordering, which is 
static within the NMR time scale.  

\begin{figure}
\includegraphics[width=8.5cm, height=7cm]{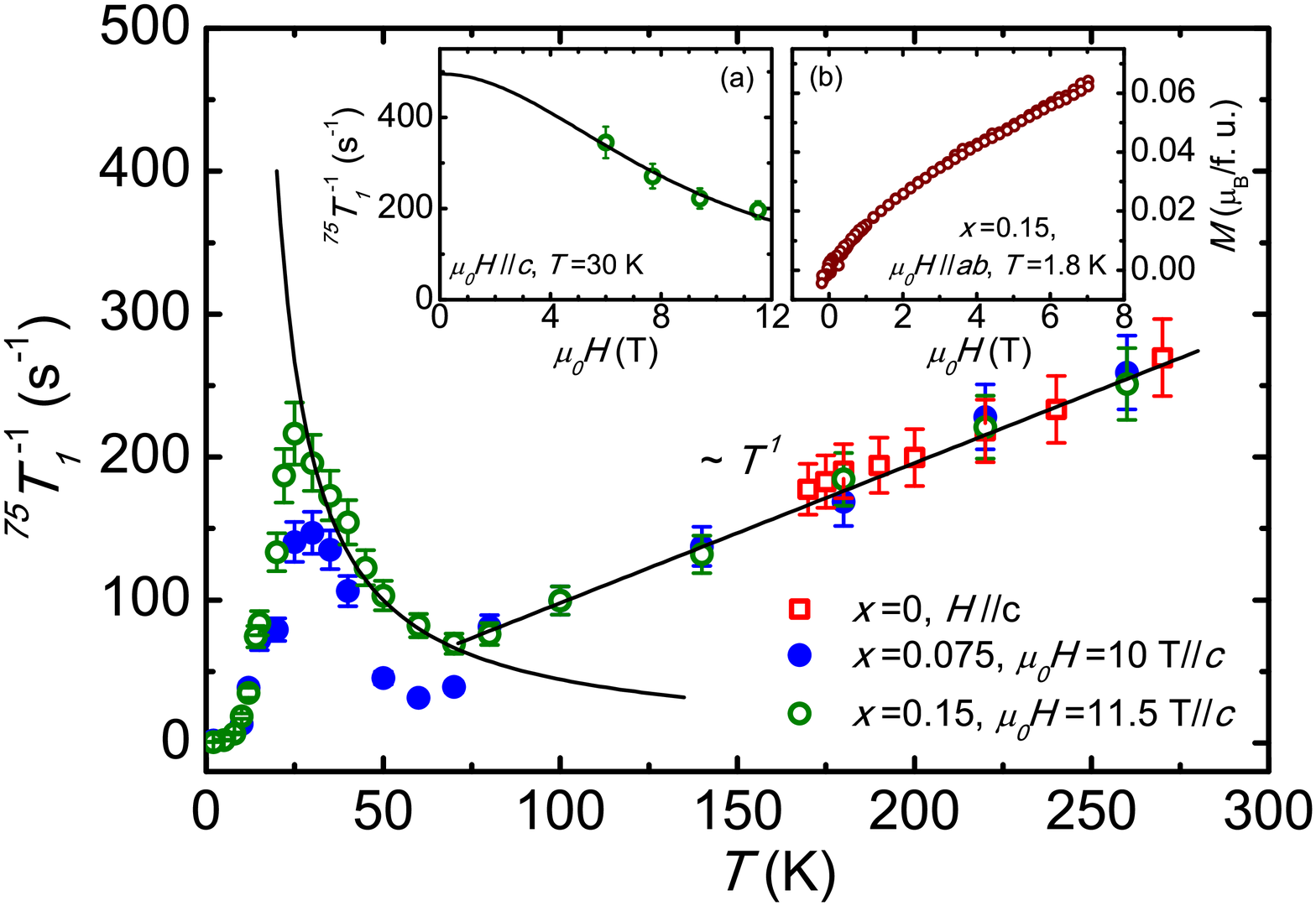}
\caption{\label{slrr}(color online) Main panel: The $^{75}$As spin-lattice relaxation rate ($1/^{75}T_1$) of Ca$_{1-x}$Pr$_x$Fe$_2$As$_2$
as a function of temperature with $x$$=$0 (adapted from Ref.~\onlinecite{Baek_PRB_79_052504}), 0.075, and 0.15. The curved solid line is a fit to the Curie-Weiss function (see text). (a): The 
spin-lattice relaxation rates of the $x$$=$0.15 sample as a function of field at $T$=30 K. The black solid line is a guide for the eye. Inset(b): The dc magnetization of the $x$$=$0.15 sample as a 
function of field applied along the $ab$ plane.}
\end{figure}

We now search for evidence of magnetic ordering from the spin-lattice relaxation rate $1/^{75}T_1$. In Fig.~\ref{slrr}, the temperature dependence of the $1/^{75}T_1$ are shown for undoped 
CaFe$_2$As$_2$ and our two dopings. Above $T_S$, $1/^{75}T_1$ is proportional to the temperature for all dopings, as shown by the straight line. Below $T_S$, however, a large increase of 
$1/^{75}T_{1}$ is clearly seen for the doped samples. At $T^*\approx$ 25 K, $1/^{75}T_{1}$ shows a broad peak behavior, with $1/^{75}T_{1}$ enhanced by a factor of three comparing with the data at 80 
K. With further cooling below $T^*$, $1/^{75}T_{1}$ decreases again. In fact, the $1/T_1$ above $T^*$ can be fit by a Curie-Weiss function with $1/T_1T=A/(T+\theta)+b$, as shown by a solid line in the 
main panel of Fig.~\ref{slrr}. The peak and the Curie-Weiss behavior in $1/^{75}T_1$ clearly indicate a critical slowing down behavior of magnetic ordering, consistent with Moriya's spin fluctuation 
theory of itinerant magnet \cite{Moriya}.  
 
The reduced $^{75}K(T)$ and the increased $1/^{75}T_1$ just below $T_S$ suggests these two have very different magnetic origins. We believe that the low-temperature magnetism originates from 
Pr$^{3+}$, and not from Fe for several reasons. i) In PrFeAsO \cite{Kimber_PRB_78_140503, Zhao_PRB_78_132504} and PrFeAsO$_{1-x}$F$_x$\cite {Yamashita_physicaC_470_375}, short-range ordering of 
Pr$^{3+}$ has been reported and produces a peak behavior in $1/^{75}T_1$ at the ordering temperature $T^*$$\approx$ 10 K. The magnetic interactions of Pr$^{3+}$, distant from the FeAs plane, is likely 
mediated by the Ruderman-Kittel-Kasuya-Yoshida (RKKY) interactions from itinerant electrons. The higher ordering temperature ($T^*$) of Ca$_x$Pr$_{1-x}$Fe$_2$As$_2$ is probably caused by stronger RKKY 
interaction with larger carrier densities and/or smaller $c$-axis lattice parameters in the $c\mathcal{T}$ phase. ii) The low-temperature $1/^{75}T_1$ of Ca$_x$Pr$_{1-x}$Fe$_2$As$_2$ is strongly 
suppressed by external field, which is typical for weakly magnetic metals, as proposed by K. Ueda \cite{Ueda_NMR}. In the inset of Fig.~\ref{slrr}, the the $1/T_1$ under different fields are shown at 
$T=$30 K for the $x$=0.15 doping, where the $1/^{75}T_1$ decreases by 1.5 times with field increases from 6 T to 11.5 T. The field suppressed $1/^{75}T_1$ suggests that either the magnetic moment is 
small, or the magnetism is not strongly coupled to itinerant electrons, which is unlikely to be from the Fe moment. iii) The dc susceptibility for our two samples, as shown in the inset of 
Fig.~\ref{MT}(b) and (c), has a large low-temperature upturn and increases with doping, consistent with increasing Pr$^{3+}$ moment with doping. In contrast, Ca$_{0.85}$La$_{0.15}$Fe$_2$As$_2$ does 
not has such an large upturn \cite{Saha_arxiv_1105_4798}, which is reasonable because La$^{3+}$ is non-magnetic. 

We further propose that the Pr$^{3+}$ magnetic ordering is antiferromagnetic and short-ranged. First, the absence of increase in $^{75}K(T)$ with reducing temperatures suggests that the magnetic 
ordering is antiferromagnetic, since the Knight shift $K(T)$ measures the susceptibility $\chi(q)$ at $q=0$, whereas $1/T_{1}$($\sim$${\sum _q}A^2_{hf}\frac{Im\chi^{+-}(q,f)}{f}$) measures the 
summation of weighted contribution of $\chi(q)$ from all momentum. Second, the low-temperature susceptibility data, as shown in the inset of Fig.~\ref{MT}(b) and (c), follows a Curie-Weiss behavior 
with $\chi(T)$$\sim$$1/(T+\theta)$ ($\theta$$\approx$-25 K) and does not saturate with temperature down to 2 K. The magnetization at $T$$=$1.8 K, as shown in Fig.~\ref{slrr}(b), also increases 
linearly with field up to 7 T. Such temperature and field behavior of magnetization is inconsistent with long-range magnetic ordering of rare earths.    

Therefore, the Knight shift, the linewidth, and the spin-lattice relaxation rate data suggest that the system has a short-range antiferromagnetic ordering from Pr$^{3+}$, whereas the paramagnetic 
fluctuations from Fe is strongly suppressed in the $c\mathcal{T}$ phase. To our knowledge, this is the first experimental evidence for reduced local moment of Fe in the $c\mathcal{T}$ phase. This also 
clarifies that the paramagnetism is quenched in the $c\mathcal{T}$ phase \cite{Yildirim_PRL_102_037003} and not in the $\mathcal{T}$ phase \cite{Ji_PRB_83_132504} in our case, and expands the INS 
study beyond the AFM wave vector \cite{Pratt_prb_79_060510}. Furthermore, the suppressed Fe moments below $T_S$ suggest that interlayer distance strongly affects the magnetism, which opens a new 
avenue for studying the impact of structure effect, such as the As-Fe-As bond angle, to the magnetism and superconductivity as well.

To summarize, we have carried out $^{75}$As NMR study on the rare-earth doped CaFe$_2$As$_2$ for the first time. For Ca$_{1-x}$Pr$_{x}$Fe$_2$As$_2$ ($x=$0.075, 0.15) superconducting single crystals, 
the chemical pressure effect by doping and the structure collapse are identified from NMR. A large drop of the Knight shift right below the structural collapse suggest a large suppression of local 
paramagnetism from Fe. Evidence for critical slowing down behavior is found in the spin-lattice relaxation at $T^*$ $\sim$ 25 K, which is attributed to short-range antiferromagnetic ordering of 
Pr$^{3+}$. Our results are not only important for understanding the interplay between structure and magnetism, but also point out that further study on the superconducting phase of the rare-earth 
doped CaFe$_2$As$_2$ should also be important to reveal the nature of superconductivity in iron pnictides in general.    

Work at RUC is supported by the NSFC (Grant No. 11074304) and the National Basic Research Program of China (Grant No. 2010CB923004 and 2011CBA00112). Work at UMD is supported by AFOSR-MURI Grant No. 
FA9550-09-1-0603.

\end{document}